\documentclass{emulateapj}
\usepackage{amsmath}
\usepackage{graphicx}

\shorttitle{Electron Acceleration by Plasma Instabilities in Solar Flares}
\shortauthors{}

\begin{document}

\title{Stochastic Electron Acceleration by Temperature Anisotropy Instabilities \\Under Solar Flare Plasma Conditions}  
\author{Mario Riquelme\altaffilmark{1}, Alvaro Osorio\altaffilmark{1}, Daniel Verscharen\altaffilmark{2,}\altaffilmark{3}, Lorenzo Sironi\altaffilmark{4}}
\altaffiltext{1}{Departamento de F\'isica, Facultad de Ciencias F\'isicas y Matem\'aticas, Universidad de Chile}
\altaffiltext{2}{Mullard Space Science Laboratory, University College London, Dorking, Surrey, UK}
\altaffiltext{3}{Space Science Center, University of New Hampshire, Durham, NH, USA}
\altaffiltext{4}{Department of Astronomy, Columbia University, New York, NY 10027, USA}

\begin{abstract} 
\noindent  Using 2D particle-in-cell (PIC) plasma simulations we study electron acceleration by temperature anisotropy instabilities, assuming conditions typical of above-the-loop-top (ALT) sources in solar flares. We focus on the long-term effect of $T_{e,\perp} > T_{e,\parallel}$ instabilities by driving the anisotropy growth during the entire simulation time, through imposing a shearing or a compressing plasma velocity ($T_{e,\perp}$ and $T_{e,\parallel}$ are the temperatures perpendicular and parallel to the magnetic field). This magnetic growth makes $T_{e,\perp}/T_{e,\parallel}$ grow due to electron magnetic moment conservation, and amplifies the ratio $\omega_{ce}/\omega_{pe}$ from $\sim 0.53$ to $\sim 2$ ($\omega_{ce}$ and $\omega_{pe}$ are the electron cyclotron and plasma frequencies, respectively). In the regime $\omega_{ce}/\omega_{pe}\lesssim 1.2-1.7$ the instability is dominated by oblique, quasi-electrostatic (OQES) modes, and the acceleration is inefficient. When $\omega_{ce}/\omega_{pe}$ has grown to $\omega_{ce}/\omega_{pe}\gtrsim 1.2-1.7$, electrons are efficiently accelerated by the inelastic scattering provided by unstable parallel, electromagnetic z (PEMZ) modes. After $\omega_{ce}/\omega_{pe}$ reaches $\sim 2$, the electron energy spectra show nonthermal tails that differ between the shearing and compressing cases. In the shearing case, the tail resembles a power-law of index $\alpha_s \sim$ 2.9 plus a high-energy bump reaching $\sim 300$ keV. In the compressing runs, $\alpha_s \sim$ 3.7 with a spectral break above $\sim 500$ keV. This difference can be explained by the different temperature evolutions in these two types of simulations, suggesting a critical role played by the type of anisotropy driving, $\omega_{ce}/\omega_{pe}$ and the electron temperature in the efficiency of the acceleration.
\end{abstract}  


\section{Introduction}
\label{sec:intro}

\noindent The mechanism responsible for electron acceleration in solar flares is a longstanding open problem in solar physics \citep[see][for reviews]{MillerEtAl1997,BenzEtAl2010,FletcherEtAl2011,Benz2017,OkaEtAl2018,Dahlin2020,LiEtAl2021}. Magnetic reconnection within prominent magnetic loops in the solar corona is thought to be the primary process for converting magnetic energy into kinetic plasma energy \citep{CairnsEtAl2018,WangEtAl2017,GouEtAl2017,LiEtAl2017a,ZhuEtAl2016,Forbes2013,Miller2013,SuEtAl2013}. However, the dominant mechanism by which electrons are accelerated and the conditions under which it operates remain to be fully understood. \newline

\noindent The leading candidate for the location of the acceleration is the reconnection current sheets (CS), where multiple processes are expected to be relevant, including convective and magnetic-field-aligned electric fields \citep{Kliem1994, DrakeEtAl2005, EgedalEtAl2012, WangEtAl2016} and Fermi-type reflections in coalescencing and contracting plasmoids formed by the tearing mode instability of the CS \citep{DrakeEtAl2006,DrakeEtAl2013,LerouxEtAl2015,DuEtAl2018}. Recent kinetic simulations are showing that 
electron spectra with nonthermal, power-law tails consistent with observations can be produced within the reconnecting CS \citep{LiEtAl2019,CheEtAl2020,CheEtAl2021, ArnoldEtAl2021, ZhangEtAl2021}, and the role played by plasma conditions in this acceleration (e.g., plasma $\beta$ and guide field strength) is currently under investigation \citep[see][for recent discussions on the open questions]{Dahlin2020,LiEtAl2021}.\newline

\noindent Additionally, there is observational evidence suggesting that part of the electron acceleration can also occur outside the CS, in a so called above-the-loop-top (ALT) region, located between the bottom of the CS and the top of the magnetic loops \citep[e.g.,][]{LiuEtAl2008,ChenEtAl2020, ChenEtAl2021}. This region constitutes a highly dynamic environment where a significant fraction of the energy carried by the reconnection outflow is dissipated, opening the possibility for several electron acceleration processes to occur. For instance, 
as the reconnection outflows impinge upon the top of the flare loops, a termination shock (TS) can form \citep{ChenEtAl2019, LuoEtAl2021}, potentially giving rise to efficient diffusive shock acceleration \citep{ChenEtAl2015}. Also, magnetohydrodynamic (MHD) simulations show that the plasmoids formed in the reconnection CS can generate a highly turbulent TS downstream medium \citep{CaiEtAl2019,KongEtAl2020,TakasaoEtAl2015,ShenEtAl2018}. This turbulent environment has been considered as a possible site for efficient stochastic electron acceleration driven by various plasma waves, 
including fast magnetosonic waves \citep{MillerEtAl1996,Miller1997,PongkitiwanichakulEtAl2014} and whistler waves \citep{HamiltonEtAl1992,PetrosianEtAl2004}. In these models, the waves are generated by MHD turbulence cascade, and their acceleration efficiencies rely on various assumptions, such as the amplitude and spectral energy distribution of the relevant modes \citep{Petrosian2012,KleinEtAl2017}. \newline

\noindent In addition to being driven by the turbulent cascade, it has been proposed that stochastic acceleration may also be due to waves excited by electron temperature anisotropy instabilities \citep{MelroseEtAl1974}. In this scenario, the temperature anisotropy may be caused by local variations of the turbulence magnetic field, which can make the electron distribution anisotropic due to the adiabatic invariance of the electron magnetic moment $\mu_e$ ($\propto v_{e,\perp}^2/\textit{B}$, where $v_{e,\perp}$ is the electron velocity perpendicular to $\textbf{\textit{B}}$).\newline

\noindent In this work, we build upon this idea and use 2D particle-in-cell (PIC) simulations to study the possible role of electron anisotropy instabilities for stochastically accelerating electrons in ALT regions. We consider the case where the temperature perpendicular to the magnetic field $\textbf{\textit{B}}$ ($T_{e,\perp}$) is larger than the parallel temperature ($T_{e,\parallel}$). Besides the possibility of this anisotropy being produced by local magnetic field growth due to turbulence, $T_{e,\perp} > T_{e,\parallel}$ can also be due to the fact that ALT regions can act as magnetic traps. Indeed, the increase of the magnetic field of the loops towards the solar surface should produce a magnetic mirror that traps large pitch-angle electrons, probably forming an anisotropic `loss-cone' electron velocity distribution with $T_{e,\perp} > T_{e,\parallel}$ \citep[e.g.,][]{FleishmanEtAl1998}. In addition, these traps are expected to behave as ``collapsing traps" as newly reconnected magnetic field lines tend to pileup on the top of the magnetic loops, producing an overall growth of the magnetic field within them. 
This magnetic growth may also contribute to the increase of the $T_{e,\perp} > T_{e,\parallel}$ anisotropy, as it has been shown by previous test particle studies of electron evolution in collapsing traps \citep{KarlickyEtAl2004,MinoshimaEtAl2010,XiaEtAl2020}. Despite these considerations, to date there is no direct observational evidence of a $T_{e,\perp} > T_{e,\parallel}$ temperature anisotropy in ALT regions and, therefore, the generation of this anisotropy is an assumption in our work.\newline

\noindent Our study is in part motivated by previous PIC simulation studies of temperature anisotropy instabilities in regimes similar to solar flares, where nonthermal electron acceleration has been found \citep{GaryEtAl2011,ChangEtAl2013,TaoEtAl2014, AnEtAl2017,LeeEtAl2018,AbdulEtAl2021}. In these studies, arbitrary values for the initial $\Delta T_e/T_{e,\parallel}$ ($\equiv (T_{e,\perp} - T_{e,\parallel})/T_{e,\parallel}$) are imposed in the simulations, with the chosen value of the anisotropy playing a critical role in determining the efficiency of the acceleration \citep[see, e.g.,][where the generation of kappa distributions is found, with the $\kappa$-parameter depending on $\Delta T_e/T_{e,\parallel}$]{TaoEtAl2014}. In our simulations we adopt a different approach, by driving the anisotropy through an (externally imposed) magnetic field growth. Including this driving is important because it allows $\Delta T_e/T_{e,\parallel}$ to be limited by the anisotropy threshold for the growth of the unstable modes, which is an important aspect in the evolution of $\Delta T_e/T_{e,\parallel}$ in real systems. Indeed, the existence of these thresholds has been predicted by plasma kinetic theory \citep[e.g.,][]{GaryEtAl1999} and it has been verified by previous PIC simulations \citep[e.g.,][]{SironiEtAl2015,Sironi2015,RiquelmeEtAl2015,RiquelmeEtAl2016,RiquelmeEtAl2017,RiquelmeEtAl2018}, and by in-situ measurements in the solar wind \citep[e.g.,][]{StverakEtAl2008}. In addition, including the anisotropy driving allows the simulations to capture in a self-consistent way the long term effects of the unstable modes on the nonthermal component of the electron velocity distribution and vice versa, as it has been shown by previous PIC studies regarding semirelativistic plasmas relevant for hot accretion flows around black holes \citep[][]{RiquelmeEtAl2017,LeyEtAl2019}.\newline


\noindent Our approach thus is to drive the growth of a $T_{e,\perp} > T_{e,\parallel}$ anisotropy by externally imposing a macroscopic plasma motion that continuously amplifies the local magnetic field due to magnetic flux freezing. After the anisotropy reaches the threshold for the growth of the unstable modes, these modes pitch-angle scatter the particles, maintaining the anisotropy at a self-regulated level and modifying the electron velocity distribution. One of our goals is to understand the sensitivity of the electron acceleration to the type of anisotropy driving. Thus in our runs we force the growth of the field by imposing either a shearing or a compressing plasma motion (hereafter, shearing and compressing simulations, respectively). We show below that the acceleration efficiencies obtained from these two driving strategies are significantly different.\newline

\noindent Some relevant considerations regarding our simulation strategy are: $i)$ Although our simulations include slow (MHD-like) bulk plasma velocities, the simulation domains are much smaller than typical MHD length scales. This way, our runs focus on the microphysics of the interaction between electrons and the unstable plasma modes by zooming in on the kinetic length scales of the modes (typically close to the electron Larmor radius $R_{Le}$), and taking the MHD evolution as an external driver. $ii)$ Our simulations use homogeneous domains with periodic boundary conditions. This means that we do not account for the loss of small pitch-angle electrons, as expected from magnetic trap configurations, therefore ignoring the possible formation of a loss-cone velocity distribution. $iii)$ In order to optimize our computational resources, we assume infinitely massive ions. This way only the electron-scale dynamics is captured, with the immobile ions only providing a neutralizing charge density to the plasma.\footnote{As it was done in a previous study of electron acceleration by temperature anisotropy-driven instabilities in the context of semirelativistic plasmas, relevant for accretion flows around black holes (Riquelme et al 2017).}\newline

\noindent This manuscript is organized as follows. In \S \ref{sec:numsetup} we describe our simulations setup. In \S \ref{sec:visheating} we use shearing simulations to show how the instabilities regulate electron temperature anisotropy. In \S \ref{enta} we use shearing simulations to show the way the instabilities produce nonthermal electron acceleration. In \S \ref{sec:compressing} we show the compressing case, emphasizing the differences and similarities with the shearing runs. In \S \ref{sec:collisions} we briefly discuss the possible role of Coulomb collisions. Finally, in \S \ref{sec:conclu} we present our conclusions.

\section{Simulation Setup}
 \label{sec:numsetup}
 \begin{deluxetable}{llllll} \tablecaption{{\bf Shearing Simulations Parameters}} \tablehead{ \colhead{Run}&\colhead{$\omega_{ce}^{\textrm{init}}/s$}&\colhead{N$_{\textrm{epc}}$}&\colhead{$d_{e}^{\textrm{init}}/\Delta_x$}&\colhead{L/R$_{Le}^{\textrm{init}}$} & $c/[\Delta_x/\Delta_t]$} 
\startdata
  S300  & 300& 100 & 35&140&0.225\\
  S600  & 600& 100 & 35&140&0.225\\  
  S1200  & 1200& 100& 35& 140&0.225\\
  S2400  & 2400& 100& 35& 140&0.225\\
  S1200a  & 1200& 100& 25& 140&0.225\\
  S1200b  & 1200& 50& 35& 140&0.225\\
  S1200c  & 1200& 100& 35& 70&0.225
\enddata 
\tablecomments{Simulation parameters for the shearing runs: the electron magnetization $\omega_{ce}^{\textrm{init}}/s$,  the number of macro-electrons per cell (N$_{\textrm{epc}}$), the initial electron skin depth $d_e^{\textrm{init}}$ ($\equiv c/\omega_{pe}^{\textrm{init}}$) in terms of grid point spacing $\Delta_x$, the box size in terms of the initial electron Larmor radius ($L/R_{Le}^{\textrm{init}}$), and the speed of light ($c/[\Delta_x/\Delta_t]$), where $\Delta_t$ is the simulation time step.} \label{table26} 
\end{deluxetable}

 \begin{deluxetable}{llllll} \tablecaption{{\bf Compressing Simulations Parameters}} \tablehead{ \colhead{Run}&\colhead{$\omega_{ce}^{\textrm{init}}/q$}&\colhead{N$_{\textrm{epc}}$}&\colhead{$d_{e}^{\textrm{init}}/\Delta_x$}&\colhead{L/R$_{Le}^{\textrm{init}}$} & $c/[\Delta_x/\Delta_t]$} 
\startdata
  C300  & 300& 200 & 50&78& 0.13\\
  C600  & 600& 200 & 50&78& 0.13\\  
  C1200  & 1200& 200& 50& 78& 0.13\\
  C2400  & 2400& 200& 50& 78& 0.13\\
  C2400a  & 2400& 200& 40& 78& 0.13\\
  C2400b  & 2400& 100& 50& 78& 0.13\\
  C2400c  & 2400& 200& 50& 39& 0.13
\enddata 
\tablecomments{Same as Table \ref{table26} but for the compressing runs.} 
\label{table27} 
\end{deluxetable}

\noindent We use the particle-in-cell (PIC) code TRISTAN-MP \citep{Buneman93, Spitkovsky05}. Our 2D simulation boxes consist of an initially square-shaped domain in the $x$-$y$ plane, which initially contains a homogeneous plasma with an isotropic Maxwell-Boltzmann velocity distribution, in presence of an initial magnetic field $\textbf{\textit{B}}_0=B_0 \hat{x}$. The magnetic field is then amplified by imposing either a shearing or a compressing bulk motion in the particles, which drives the electron temperature anisotropic with $T_{\perp,e} > T_{||,e}$ due to $\mu_e$ conservation. In the shearing case, the plasma velocity is given by $\textbf{\textit{v}} = -sx\hat{y}$, where $x$ is the distance along $\hat{x}$ and $s$ is the constant shear rate (this setup is shown in Figure 1 of \cite{LeyEtAl2019}).\footnote{The shear simulations are performed in the `shearing coordinate system', in which the shearing velocity of the plasma vanishes, and both Maxwell's equations and the Lorentz force on the particles are modified accordingly \cite[see][]{RiquelmeEtAl2012}.} From flux conservation, the $y$-component of the mean field evolves as a function of time $t$ as $\langle B_y\rangle = -sB_0t$ (throughout this paper, $\langle \rangle$ represents an average over the simulation domain), implying that $|\langle \textbf{\textit{B}}\rangle |$ grows as $|\langle \textbf{\textit{B}}\rangle | = B_0(1+(st)^2)^{1/2}$.  In the compressing case, $\textbf{\textit{v}} = -q(y\hat{y}+z\hat{z})/(1+qt)$, where $q$ is a constant that quantifies the compression rate of the simulation box. In this case, $|\langle \textbf{\textit{B}}\rangle |= B_0(1+qt)^2$ \citep[the compressing setup is shown in Figure 1 of][]{SironiEtAl2015}.\newline

\begin{figure*}[t!]  
\centering 
\hspace*{-.1cm}\includegraphics[width=18.4cm]{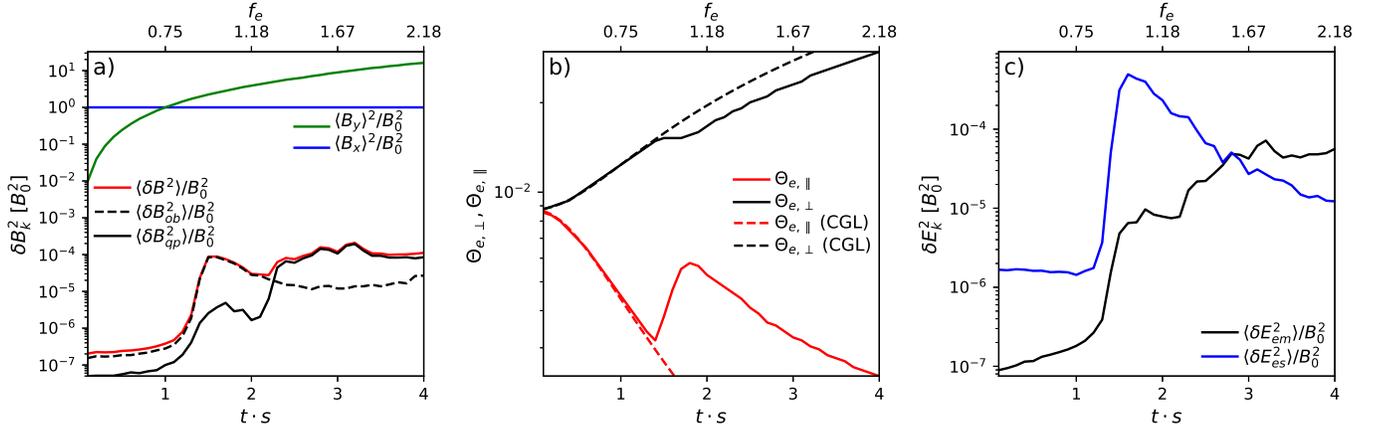}
\caption{Fields and electron temperature evolutions for run S1200 as a function of time $t$ in units of $s^{-1}$ (lower horizontal axes) and of the instantaneous $f_e$ (upper horizontal axes; using the average magnetic field at each time). Panel $\it{a}$ shows in solid-blue and solid-green lines the evolution of the energy in the $x$  and $y$ components of the mean magnetic field $\langle \textbf{\textit{B}}\rangle$, respectively. The solid-red line shows the energy in $\delta \textbf{\textit{B}}$, while the solid-black (dashed-black) line shows the contribution to the $\delta \textbf{\textit{B}}$ energy given by the quasi-parallel (oblique) modes  [all in units of the initial magnetic energy]. Panel $\it{b}$ shows in solid-black (solid-red) the evolution of the electron temperature perpendicular (parallel) to $\langle \textbf{\textit{B}}\rangle$. The dashed-black (dashed-red) line shows the CGL prediction for the perpendicular (parallel) temperature. Panel $\it{c}$ shows in black (blue) the energy in the electromagnetic and electrostatic component of the electric field fluctuations $\delta \textbf{\textit{E}}$, which satisfy $\nabla \cdot\delta \textbf{\textit{E}}=0$ and $\nabla \times \delta \textbf{\textit{E}}=0$, respectively.}
\label{fig:sf1} 
\end{figure*}

\noindent The initial plasma parameters in our simulations are chosen so that they represent typical ALT conditions. In these environments,
 reported electron temperatures are usually of a few tens MK \citep{FeldmanEtAl1994,MasudaEtAl1994,MasudaEtAl1995,FletcherEtAl2011}. Our simulations thus use an initial temperature of $52$ MK, which, when normalized by the electron rest mass energy, gives $\Theta_e^{\textrm{init}}\equiv k_B T_e^{\textrm{init}}/m_ec^2=0.00875$ ($k_B$ is the Boltzmann constant, $m_e$ is the electron mass and $c$ is the speed of light). Additionally, the magnetic field intensity, \textit{B}, is typically close to $\sim 100$ G \citep{KuridzeEtAl2019}, while the electron density, $n_e$, is usually estimated in the range $\sim 10^8-10^{12}$ cm$^{-3}$ \citep{FeldmanEtAl1994,MasudaEtAl1994,MasudaEtAl1995, TsunetaEtAl1997}. If we define the ratio
\begin{equation}
    f_e\equiv \frac{\omega_{ce}}{\omega_{pe}},
\end{equation}
where $\omega_{ce}$ and $\omega_{pe}$ are the electron cyclotron and plasma frequencies, respectively, we find that when choosing the fiducial values $\textit{B} \sim 100$ G and $n_e \sim 10^9$ cm$^{-3}$, $f_e$ becomes $\sim 1$ ($\omega_{ce}= |e|B/m_ec$ and $\omega_{pe}= (4\pi n_e e^2/m_e)^{1/2}$, where $e$ is the electron charge). Our runs thus use an initial $f_e \approx 0.53$ (implying an initial electron beta $\beta_e^{\textrm{init}}=0.0625$, where $\beta_e\equiv 8\pi n_ek_B T_e/B^2$). We run both our shearing and compressing simulations until $f_e$ has been amplified to $f_e \approx 2$. This allows us to compare the shearing and the compressing run under similar plasma conditions. This also allows us to emphasize the important role played by $f_e$ in determining both the dominant unstable modes and the efficiency of the acceleration.\newline

\noindent Another important physical parameter in our runs is the initial electron ``magnetization", defined as the ratio between the initial electron cyclotron frequency and either the shear rate ($\omega_{ce}^{\textrm{init}}/s$) or the compression rate ($\omega_{ce}^{\textrm{init}}/q$). Although the magnetizations in our runs are much larger than unity, for computational convenience we chose them much smaller than expected in real flare conditions. Thus, we ensure that our magnetizations are large enough to not affect significantly our results by using $\omega_{ce}^{\textrm{init}}/s$ (and $\omega_{ce}^{\textrm{init}}/q$) $=300, 600, 1200$ and 2400 to show that our results tend to converge as the magnetization grows. \newline

\noindent The numerical parameters in our runs are: the number of macro-electrons per cell (N$_{\textrm{epc}}$), the initial electron skin depth $d_e^{\textrm{init}}$ ($\equiv c/\omega_{pe}^{\textrm{init}}$) in terms of the grid point spacing $\Delta_x$, the initial box size $L$ in terms of the initial electron Larmor radius $R_{Le}^{\textrm{init}}$ ($\equiv v_{th,e}/\omega_{ce}^{\textrm{init}}$, where $v_{th,e}^2=k_BT_e^{\textrm{init}}/m_e$), and the speed of light $c$ in units of $\Delta_x/\Delta_t$, where $\Delta_t$ is the simulation time step. We ran a series of simulations to make sure that our choices for the magnetization and for the numerical parameters do not affect our results; these simulations are summarized in Tables \ref{table26} and \ref{table27} for the shearing and compressing runs, respectively.\newline

\begin{figure*}[t!]  
\centering 
\hspace*{-.2cm}\includegraphics[width=18.5cm]{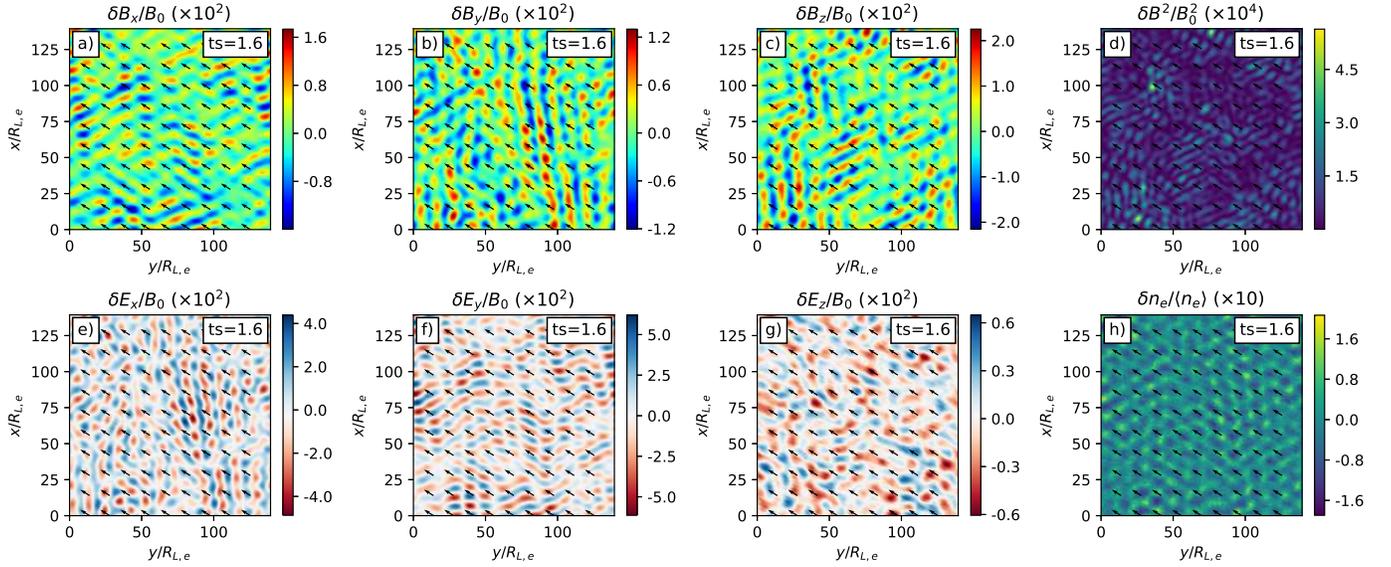}
\caption{For run S1200 we show the 2D distribution of the three components of $\delta \textbf{\textit{B}}$: $\delta \textit{B}_x$, $\delta \textit{B}_y$ and $\delta \textit{B}_z$, the three components of $\delta \textbf{\textit{E}}$: $\delta \textit{E}_x$, $\delta \textit{E}_y$ and $\delta \textit{E}_z$, the total $\delta \textbf{\textit{B}}$ energy and the electron density fluctuations $\delta n_e$ ($\equiv n_e-\langle n_e \rangle$) at $t\cdot s = 1.6$. Fields and density are normalized by $B_0$ and by the average density $\langle n_{e} \rangle$, respectively. The black arrows show the direction of the average magnetic field $\langle \textbf{\textit{B}} \rangle$.} 
\label{fig:fldsf_ts16} 
\end{figure*}
\begin{figure*}[t!]  
\centering 
\hspace*{-.2cm}\includegraphics[width=18.5cm]{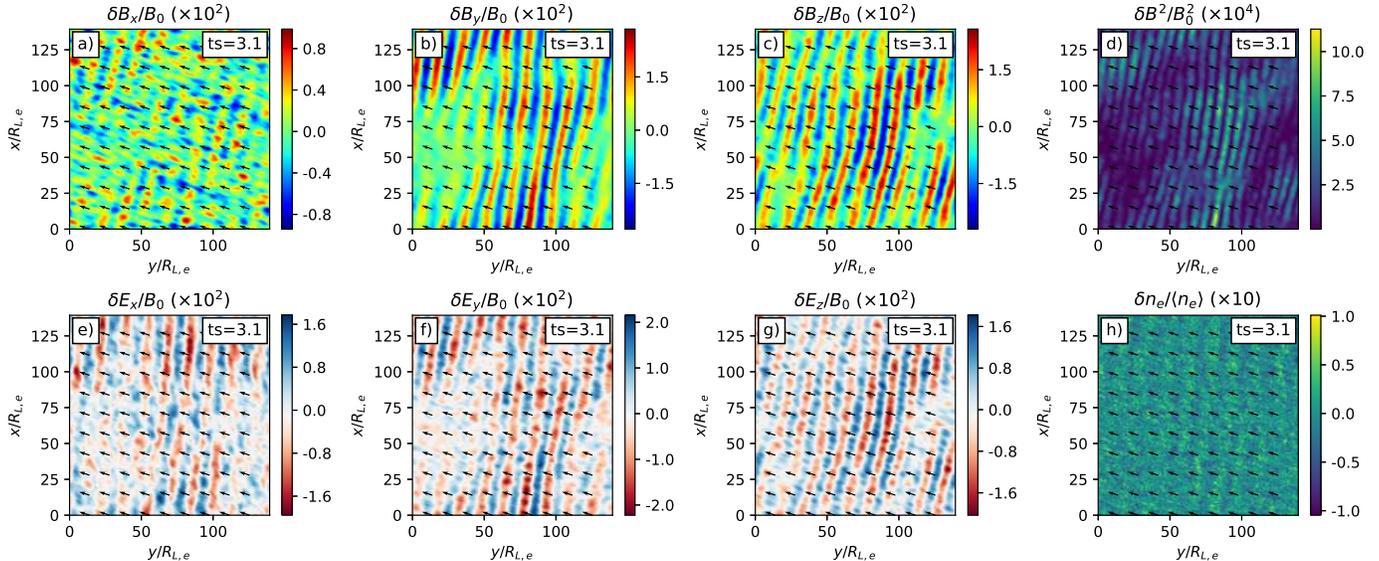}
\caption{Same as Fig. \ref{fig:fldsf_ts16} but at $t\cdot s = 3.1$.} 
\label{fig:fldsf_ts31} 
\end{figure*} 

\section{Electron temperature anisotropy regulation}
\label{sec:visheating}
\noindent Before describing the effect of unstable plasma modes in producing electron acceleration, in this section we describe the way these modes regulate the temperature anisotropy. Since this regulation is qualitatively similar in the shearing and compressing runs, our description is based on the shearing simulations.

\subsection{Interplay between magnetic field growth and temperature anisotropy evolution}
\label{interplay}
Figure \ref{fig:sf1}$a$ shows in solid-green the linear growth of the $y$ component of the mean magnetic field $\langle \textbf{\textit{B}}\rangle $ in the shearing simulation S1200 (see Table \ref{table26}). The $x$ component remains constant at a value of $B_0$ and the $z$ (out of plane) component is zero, as expected in our shearing setup. In Figure \ref{fig:sf1}$b$ we see that, due to the growth of $|\langle \textbf{\textit{B}}\rangle |$, the electron temperatures perpendicular and parallel to $\langle \textbf{\textit{B}}\rangle$, $\Theta_{e,\perp}$ ($\equiv k_BT_{e,\perp}/m_ec^2$; solid-black) and $\Theta_{e,\parallel}$ ($\equiv k_BT_{e,\parallel}/m_ec^2$; solid-red), grow and decrease, respectively, as expected from their initially adiabatic evolutions. Indeed, in the initial regime, $\Theta_{e,\perp}$ and $\Theta_{e,\parallel}$ evolve according to the adiabatic Chew-Goldberg-Low (hereafter, CGL) equation of state \citep{ChewEtAl1956}, shown by the dashed-black and dashed-red lines, respectively.\footnote{The CGL equation of state implies that $T_{e,\perp}/ B$ and $T_{e,\parallel}B^2/n_e^2$ remain constant.} The departure from the CGL evolution at $t\cdot s \approx 1.4$ coincides with the rapid growth and saturation of $\delta \textbf{\textit{B}}$ ($\equiv \textbf{\textit{B}} - \langle \textbf{\textit{B}} \rangle$), as shown by the solid-red line in Figure \ref{fig:sf1}$a$. This shows that the growth of the temperature anisotropy is ultimately limited by temperature anisotropy unstable modes, which can break the adiabatic evolution of the electron temperatures by providing efficient pitch-angle scattering.\newline 

\subsection{The nature of the unstable modes}
\label{sec:nature}
\noindent A 2D view of the relevant unstable modes at $t\cdot s = 1.6$ (right after the saturation of $\delta \textbf{\textit{B}}$) is given by Figure \ref{fig:fldsf_ts16}, which shows the three components of the magnetic and electric fluctuations $\delta \textbf{\textit{B}}$ and $\delta \textbf{\textit{E}}$.\footnote{$\delta \textbf{\textit{E}}$ is simply equal to $\textbf{\textit{E}}$, since $\langle \textbf{\textit{E}} \rangle=0$ in our shearing coordinate setup.} Both the magnetic and electric fluctuations show that the dominant modes have an oblique wavevector with respect to the direction of the mean magnetic field $\langle \textbf{\textit{B}} \rangle$, which is shown by the black arrows in all the panels. The dominance of the oblique modes stops at a later time, as can be seen from Figure \ref{fig:fldsf_ts31}, which shows the same quantities as Figure \ref{fig:fldsf_ts16} but at $t\cdot s = 3.1$. In this case the waves propagate along the background magnetic field, implying that quasi-parallel modes dominate both the electric and magnetic fluctuations.\newline

\noindent In order to determine the transition time from the dominance of oblique to quasi-parallel modes, we calculate the magnetic energy contained in each type of modes. For that, we define the modes as ``oblique" or ``quasi-parallel" depending on whether the angle between their wave vector $\textbf{\textit{k}}$ and $\langle\textbf{\textit{B}}\rangle$ is larger or smaller than 20$^{\circ}$, respectively. The magnetic energies in oblique and quasi-parallel modes are shown in Figure \ref{fig:sf1}$a$ using dashed-black and solid-black lines and are denoted by $\langle \delta B_{ob}^2 \rangle$ and $\langle \delta B_{qp}^2 \rangle$, respectively. We see that the oblique modes dominate until $t\cdot s \approx 2.2$. After that, the quasi-parallel modes contribute most of the energy of the magnetic fluctuations, which is consistent with the two regimes shown in Figures \ref{fig:fldsf_ts16} and \ref{fig:fldsf_ts31}.\newline

\noindent An interesting characteristic of the oblique modes is the notorious fluctuations in the electron density $n_e$, as shown in Figure \ref{fig:fldsf_ts16}$h$, which suggests the presence of a significant electrostatic component in the electric field $\delta \textbf{\textit{E}}$. These density fluctuations are less prominent in the case dominated by quasi-parallel modes, as can be seen from Figure \ref{fig:fldsf_ts31}$h$, which shows that at $t\cdot s=3.1$ the electrostatic electric fields are weaker than at $t\cdot s=1.6$. This change in the $\delta \textbf{\textit{E}}$ behavior can be seen more clearly in Figure \ref{fig:sf1}$c$, which shows the evolution of the energy contained in the electric field fluctuations $\delta \textbf{\textit{E}}$, dividing it into its electrostatic and electromagnetic components (blue and black lines, respectively). This separation is achieved by distinguishing the contributions to the Fourier transform of the electric field fluctuations ($\delta \tilde{\textbf{\textit{E}}}$) that satisfy  $\textbf{\textit{k}} \times \delta \tilde{\textbf{\textit{E}}} = 0$ (electrostatic part) and $\textbf{\textit{k}} \cdot \delta \tilde{\textbf{\textit{E}}} = 0$ (electromagnetic part). In the oblique regime ($t\cdot s \lesssim 2.2$) the electric field energy is mainly dominated by its electrostatic part ($\delta E^2_{es}$), and after that it gradually becomes dominated by its electromagnetic part ($\delta E^2_{em}$). Figure \ref{fig:sf1}$c$ also shows that this transition occurs when the instantaneous parameter $f_e$ (shown by the upper horizontal axis) is $f_e\sim 1.2-1.5$. We see in the next section that this transition is fairly consistent with linear Vlasov theory.\newline

\begin{figure}[t!]  
\centering 
\includegraphics[width=8.8cm]{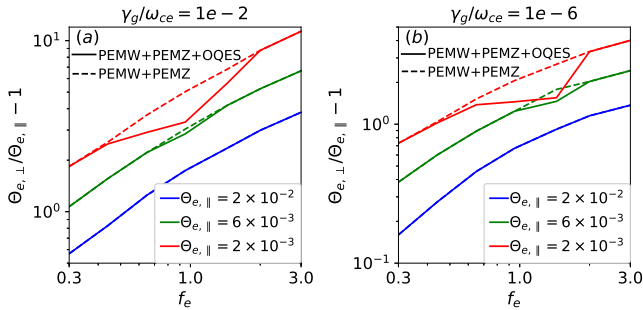}
\setlength{\abovecaptionskip}{-6pt plus 3pt minus 2pt}
\caption{The anisotropy thresholds $\Theta_{e,\perp}/\Theta_{e,\parallel}-1$ for the growth of parallel, electromagnetic PEMW and PEMZ modes (dashed lines) and for the combination of PEMW, PEMZ and OQES modes (solid lines) as a function of $f_e$ and for $\Theta_{e,\parallel}=0.002$ (red), 0.006 (green) and 0.02 (blue). Calculations were performed using the NHDS solver of \cite{VerscharenEtAl2018}. Panels $a$ and $b$ show the cases with growth rate $\gamma_g=10^{-2}\omega_{ce}$ and $\gamma_g=10^{-6}\omega_{ce}$, respectively.} 
\label{fig:anisos} 
\end{figure} 

\subsection{Comparison with linear Vlasov theory}
\label{sec:consistency}
\noindent In this section, we show that the transition from the dominance of oblique modes with mainly electrostatic electric field (hereafter, oblique electrostatic modes) to the dominance of quasi-parallel modes with mainly electromagnetic electric field (hereafter, quasi-parallel electromagnetic modes) in S1200 is consistent with linear Vlasov theory. Indeed, a previous study by \cite{GaryEtAl1999} predicts that for the range of electron conditions considered in our study, and assuming a bi-Maxwellian electron velocity distribution, three types of modes are relevant: parallel, electromagnetic whistler (PEMW) modes; parallel, electromagnetic z (PEMZ) modes;  and oblique, quasi-electrostatic (OQES) modes, where the latter are dominated by electrostatic electric fields. In this section we use linear Vlasov theory to check whether the PEMW or PEMZ (OQES) modes are theoretically the most unstable when the quasi-parallel electromagnetic (oblique electrostatic) modes dominate in our run. For this we use the NHDS solver of \cite{VerscharenEtAl2018} to calculate the temperature anisotropy threshold $\Theta_{e,\perp}/\Theta_{e,\parallel}-1$ needed for the growth of these modes with a given growth rate $\gamma_g$, assuming different values of $f_e$ and $\Theta_{e,\parallel}$.\newline

\noindent Figure \ref{fig:anisos}$a$ shows the anisotropy thresholds for $\gamma_g/\omega_{ce}=10^{-2}$, which is appropriate for run S1200. We estimate the growth rate of $\delta \textbf{\textit{B}}$ in this run from its exponential growth regime in Figure \ref{fig:sf1}$a$ ($t\cdot s \sim 1.2-1.3$), which is $\gamma_g\sim 10 s$. This implies that $\gamma_g \sim 10^{-2}\omega_{ce}^{\textrm{init}}$, given that in run S1200, $\omega_{ce}^{\textrm{init}}/s=1200$. The dashed lines in Fig. \ref{fig:anisos}$a$ consider the thresholds only for parallel modes (i.e., considering the lowest threshold between PEMW and PEMZ modes), while the solid lines consider the lowest threshold between modes with all propagation angles. We find that for $\Theta_{e,\parallel}= 0.002$ and 0.006, there are values of $f_e$ where the solid-red and solid-green lines separate from the corresponding dashed lines. These values of $f_e$, therefore, correspond to where the unstable modes are dominated by OQES modes, which, for $\Theta_{e,\parallel} = 0.002$ and 0.006, occurs when $0.4 \lesssim f_e\lesssim 1.8$  and $0.8 \lesssim f_e\lesssim 1.3$, respectively. For values of $f_e$ above and below these ranges, linear theory predicts that the most unstable modes correspond to PEMZ and PEMW modes, respectively, which is consistent with the merging of the dashed and solid lines in those regimes.\footnote{The predictions shown in Figure \ref{fig:anisos}$a$ are in good agreement with Figure 4$c$ of \cite{GaryEtAl1999}} \newline

\noindent Thus, the dominance of oblique electrostatic modes in run S1200 from the moment when the instability sets in ($f_e \sim 0.8$) until $f_e\sim 1.2-1.5$, suggests that, in order to be consistent with linear theory, $\Theta_{e,\parallel}$ should be in the range $\sim 0.002-0.006$ after the growth of the instabilities ($f_e \gtrsim 0.8$). This is indeed what is shown by Figure \ref{fig:sf1}$b$, where $\Theta_{e,\parallel}$ (solid-red line) appears in the range $\Theta_{e,\parallel}\sim 0.0025-0.005$ when $f_e \gtrsim 0.8$. \newline

\noindent These results show that the transition between the oblique, electrostatic to quasi-parallel, electromagnetic regimes at $f_e\sim 1.2-1.5$ in run S1200 is consistent with linear Vlasov theory, which predicts a transition from OQES to PEMZ modes at $f_e \sim 1.3-1.8$. Therefore, hereafter we refer to the $f_e\lesssim 1.2-1.5$ and $f_e\gtrsim 1.2-1.5$ regimes as OQES dominated and PEMZ dominated regimes, respectively.

\begin{figure*}[t!]  
\centering 
\includegraphics[width=18cm]{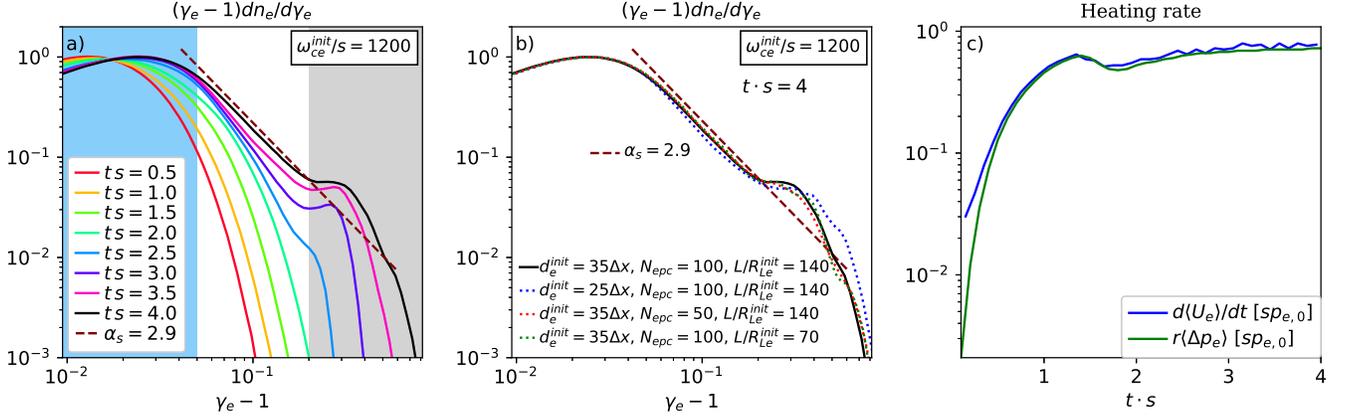}
\setlength{\abovecaptionskip}{-8pt plus 3pt minus 2pt}
\caption{\textit{Panel a:} the electron energy spectrum for run S1200 for different values of $t\cdot s$, where $\gamma_e$ is the Lorentz factor of the electrons. The dashed-brown line shows a power-law of index $\alpha_s \approx 2.9$, which resembles part of the final nonthermal tail. \textit{Panel b:} test of numerical convergence of the final spectrum. We show the electron energy spectra at $t\cdot s=4$ for runs analogous to S1200 (solid black), but using $i)$ a smaller time and space resolution $d_e^{\textrm{init}}/\Delta x=25$ (run S1200a, in blue-dotted line), $ii)$ a smaller $N_{epc}=50$ (run S1200b, in red-dotted line) and $iii)$ a smaller box size $L/R_{Le}^{\textrm{init}}=70$ (run S1200c, in green-dotted line). No significant difference can be seen between the different spectra. \textit{Panel c:} the time evolution of $d\langle U_e \rangle /dt$ (blue) and of $r \langle \Delta p_e \rangle$ (green) for the same run, normalized by $sp_{e,0}$, where $p_{e,0}$ is the initial electron pressure.} 
\label{fig:sf4} 
\end{figure*}
\section{Electron nonthermal acceleration} 
\label{enta}
\noindent Previous works show that pitch-angle scattering by temperature anisotropy instabilities can produce significant stochastic particle acceleration \citep{RiquelmeEtAl2017, LeyEtAl2019}. In this section we show that electron temperature anisotropy instabilities can also contribute significantly to the acceleration of electrons under the conditions expected in ALT regions in solar flares, which are defined by the magnetic field strength and the electron density and temperature. We show this first for the $\omega_{ce}^{\textrm{init}}/s=1200$ run S1200, and then show that the acceleration is fairly independent of $\omega_{ce}^{\textrm{init}}/s$.\newline

\subsection{Case $\omega_{ce}^{\textrm{init}}/s=1200$}
\label{case1200}
\noindent The electron energy spectrum evolution for run S1200 can be seen from Figure \ref{fig:sf4}$a$, which shows $dn_e/d\ln(\gamma_e-1)$ for different values of $t\cdot s$ ($\gamma_e$ is the electron Lorentz factor). This plot shows the rapid growth of a nonthermal tail starting at $t\cdot s\approx 2.5$. After $t\cdot s\approx 3.5$ this tail can be approximated as a power-law of index $\alpha_s \approx 2.9$ ($\alpha_s \equiv d\ln(n_e)/d\ln(\gamma_e-1)$), plus a high energy bump that reaches $\gamma_e-1\sim 0.6$ ($\sim 300$ keV). Most of the nonthermal behavior of the spectrum starts at $t\cdot s \approx 2.5$, which is right after the PEMZ modes become dominant, as shown by Figure \ref{fig:sf1}$c$. \newline

\noindent Figure \ref{fig:sf4}$b$ shows a numerical convergence test of the final spectrum at $t\cdot s=4$. It compares run S1200 ($d_e^{\textrm{init}}/\Delta x=35$, $N_{epc}=100$ and $L/R_{Le}^{\textrm{init}}=140$) with a run with $d_e^{\textrm{init}}/\Delta x=25$ (run S1200a, in blue-dotted line), a run with $N_{epc}=50$ (run S1200b, in red-dotted line) and a run with $L/R_{Le}^{\textrm{init}}=70$ (run S1200c, in green-dotted line). No significant difference can be seen between the different spectra, implying that our results are fairly converged numerically.\newline

\noindent In order to identify the energy source for the nonthermal electron acceleration, we explore first the overall energy source for electrons (thermal and nonthermal). It is well known that the presence of a temperature anisotropy in a shearing, collisionless plasma gives rise to particle heating due to the so called ``anisotropic viscosity" (AV). This viscosity can give rise to an overall electron heating, for which the time derivative of the electron internal energy, $U_e$, is (Kulsrud 1983; Snyder et al. 1997): 
\begin{equation}
\frac{dU_e}{dt} = r \Delta p_e, 
\label{eq:av}
\end{equation}
where $r$ is the growth rate of the field (in our setup $r=-sB_xB_y/B^2$) and $\Delta p_e$ is the difference between the perpendicular and parallel electron pressures,  $\Delta p_e=p_{e,\perp}-p_{e,\parallel}$. In run S1200, Equation \ref{eq:av} reproduces well the evolution of the overall electron energy gain. This can be seen from Figure \ref{fig:sf4}$c$, where the time derivative of the average electron internal energy $d\langle U_e\rangle/dt$ (blue) coincides well with $r \langle \Delta p_e \rangle$ (green). This result shows that in our shearing setup, the heating by AV essentially explains all of the electron energization.\newline
\begin{figure*}[t!]  
\centering 
\includegraphics[width=18cm]{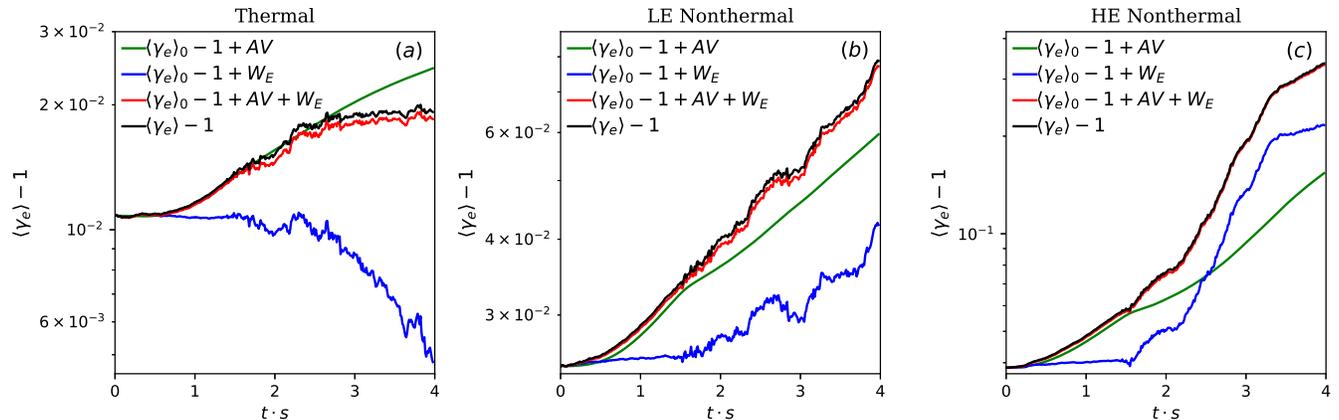}
\setlength{\abovecaptionskip}{-8pt plus 3pt minus 2pt}
\caption{The different contributions to electron energization for three electron populations in run S1200, normalized by $m_ec^2$. \textit{Panel a} shows the case of the thermal electrons, defined by their final ($t \cdot s=4$) Lorentz factor being $\gamma_e-1 <0.05$ (marked by the light-blue region of Figure \ref{fig:sf4}$a$). \textit{Panels b} and $c$ show the low-energy and high-energy nonthermal electrons, defined by their final Lorentz factor being in the ranges $0.05<\gamma_e-1 <0.2$ and  $0.2<\gamma_e-1$, respectively (marked by the white and grey regions in Figure \ref{fig:sf4}$a$, respectively). In all three panels, the black line shows the average energy evolution for the respective electron population. The green line shows their average energy gain due to anisotropic viscosity (AV). The blue lines shows the same but considering the energization by the electric field of the unstable modes ($W_E$) instead of the energization by AV. The red line shows the energy gain by the addition of AV and $W_E$, and reproduces reasonably well the total energy evolutions shown as black lines.} 
\label{fig:tplowtemp} 
\end{figure*}
\noindent Although the total electron energy gain is dominated by AV, the work done by the electric field $\delta \vec{E}$ associated to the unstable modes, $W_E (\equiv\int e\delta\vec{E}\cdot d\vec{r}$, where $\vec{r}$ is the electron position), can differ significantly between electrons from different parts of the spectrum, making $W_E$ play a key role in producing the nonthermal tail by transferring energy from the thermal to the nonthermal part of the spectrum \citep{RiquelmeEtAl2017, LeyEtAl2019}. We check this by analyzing the contributions of AV and $W_E$ to the energy gain of three different electron populations, defined by their final energy at $t\cdot s=4$. These populations are:\newline
\begin{enumerate}
\item $\textit{Thermal electrons}$: their final energy satisfies $\gamma_e-1 <0.05 $, which corresponds to the energy range marked by the light-blue background in Figure \ref{fig:sf4}$a$.
\item $\textit{Low-energy nonthermal electrons}$: their final energies satisfy $0.05 < \gamma_e-1 < 0.2$, where the nonthermal tail roughly behaves as a power-law of index $\alpha_s \approx 2.9$. This energy range is marked by the white background in Figure \ref{fig:sf4}$a$. 
\item $\textit{High-energy nonthermal electrons}$: corresponding to the high-energy bump in the spectrum, defined by $0.2 < \gamma_e-1$. This energy range is marked by the grey background in Figure \ref{fig:sf4}$a$.
\end{enumerate}
Figures \ref{fig:tplowtemp}$a$, \ref{fig:tplowtemp}$b$ and \ref{fig:tplowtemp}$c$ show the energy evolution for the thermal, low-energy nonthermal and high-energy nonthermal electrons, respectively. In each case, we show the average initial kinetic energy of each population, $\langle \gamma_e \rangle_0-1$, plus:
\begin{enumerate}
\item the work done by the electric field of the unstable modes, $W_E$, shown by the blue lines.
\item the energy gain by the anisotropic viscosity, AV, shown by the green lines.
\item the total energy gain, shown by the black lines.
\end{enumerate}
The green lines show that for the three populations there is a positive energy gain due to AV. On the other hand, the blue lines in Figure \ref{fig:tplowtemp}$a$ shows that, in the case of the thermal electrons, $W_E$ produces a decrease in the electron energies. This illustrates that the thermal electrons transfer a significant part ($\sim 50 \%$) of their initial energy to the unstable modes. However, the overall heating of the thermal electrons is positive and, by the end of the simulation, reaches a factor $\sim 2$ increase in their thermal energy. Figures \ref{fig:tplowtemp}$b$ and \ref{fig:tplowtemp}$c$ show that for the low- and high-energy nonthermal electrons there is a significant growth in the electrons energy due to $W_E$, suggesting that the unstable modes transfer energy to the nonthermal particles. For the low-energy nonthermal electrons, the AV still dominates the heating, whereas for the high-energy nonthermal electrons, AV is subdominant and most of the electron energization is due to $W_E$. The red lines in Figures \ref{fig:tplowtemp}$a$, \ref{fig:tplowtemp}$b$ and \ref{fig:tplowtemp}$c$ show the overall heating of thermal electrons due to adding AV and $W_E$. The red line reproduces reasonably well the evolution of the total energy $\langle \gamma_e \rangle -1$ (black line) of the three electron populations.\newline

\noindent The way AV and $W_E$ contribute to the energization of thermal and nonthermal electrons suggests that the formation of an electron nonthermal tail is caused by the transfer of energy from the thermal to the nonthermal electrons, which is mediated by the waves electric field. This characteristic of the formation of a nonthermal tail is in line with previous results where the acceleration of ions and electrons by temperature anisotropy instabilities was studied in nearly relativistic plasmas \citep{RiquelmeEtAl2017,LeyEtAl2019}. Also, Figures \ref{fig:tplowtemp}$a$, \ref{fig:tplowtemp}$b$ and \ref{fig:tplowtemp}$c$ show that most of the energy transfer from thermal to nonthermal electrons occurs after $t\cdot s \sim 2.2$, which corresponds to the regime dominated by PEMZ modes, implying a subdominant contribution to the acceleration by the initially dominant OQES modes.\newline

\subsection{Extrapolation to the (realistic) very high $\omega_{ce}/s$ regime}
\label{sec:realistic}
\noindent The shear parameter $s$ is a measure of the rate at which temperature anisotropy growth is driven in our simulations. We can thus estimate the corresponding parameter $s$ in the contracting looptops of solar flares as the rate at which temperature anisotropy grows in these environments. This can be obtained by estimating the inverse of the time that it takes for the contracting looptops to collapse into a more stable configuration. \newline

\noindent We thus estimate $s$ by dividing the typical Alfv\'en velocity, $v_A$, in the looptops ($v_A$ should be close to the speed at which the newly reconnected loops get ejected from their current sheet) by the typical lengthscale of the contracting looptops, $L_{\textrm{LT}}$. Using our fiducial parameters $n_e \sim 10^{9}$ cm$^{-3}$ and $B \sim 100$ G, and estimating $L_{\textrm{LT}} \sim 10^9$ cm \citep[e.g.,][]{ChenEtAl2020}, we obtain $s \sim 1$ sec$^{-1}$ and $\omega_{ce}/s\sim 10^{9}$. This value of $\omega_{ce}/s$ is several orders of magnitude larger than what can be achieved in our simulations. Therefore, two important questions arise. The first one is whether, for realistic values of $\omega_{ce}/s$, the dominance of PEMZ and OQES modes should occur for the same regimes observed in run S1200. And, if that is the case, the second question is whether the effect on electron acceleration of these modes remains the same, independently of $\omega_{ce}/s$. \newline

\begin{figure}[t!]  
\centering 
\hspace*{-.2cm}
\includegraphics[width=8.9cm]{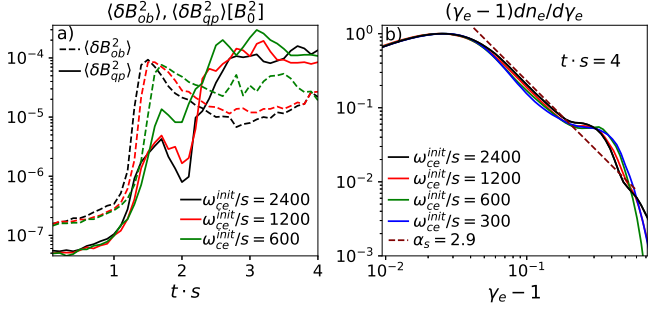}
\setlength{\abovecaptionskip}{-6pt plus 3pt minus 2pt}
\caption{Panel $a$: the evolution of the energy in $\delta \textbf{\textit{B}}$ divided into oblique (dashed) and quasi-parallel (solid) modes for runs S2400 (black), S1200 (red) and S600 (green). These runs are equal except for having $\omega_{ce}^{\textrm{init}}/s=2400$, 1200 and 600, respectively. Panel $b$: the final electron spectra for runs S2400 (black), S1200 (red), S600 (green) and S300 (blue).} 
\label{fig:fldspec} 
\end{figure}

\noindent Since in \S \ref{sec:consistency} we showed the suitability of linear Vlasov theory to predict the dominance of the different modes, in this section we use linear theory to show that increasing $\omega_{ce}^{\textrm{init}}/s$ by several orders of magnitude should not modify the relative importance of the modes in the different plasma regimes. First, we notice that the growth rate $\gamma_g$ of the unstable modes should be proportional to $s$, implying that $\omega_{ce}^{\textrm{init}}/s \propto \omega_{ce}^{\textrm{init}}/\gamma_g$. This proportionality is physically expected, since $s^{-1}$ sets the timescale for the evolution of the macroscopic plasma conditions. Thus, the different modes that dominate in different simulation stages need to have a growth rate of the order of $s$ to have time to set in and regulate the electron temperature anisotropy. This can also be seen from Figure \ref{fig:fldspec}$a$, which shows the evolution of the energy in $\delta \textbf{\textit{B}}$ divided into oblique and quasi-parallel modes (dashed and solid lines, respectively) for runs with $\omega_{ce}^{\textrm{init}}/s=600$, 1200 and 2400 (runs S600, S1200 and S2400, respectively). We see that the three runs show essentially the same ratio $\gamma_g/s\sim 10$. This means that, in order to find out which instabilities would dominate for realistically large values of $\omega_{ce}^{\textrm{init}}/s$, we must calculate the instability thresholds for a comparatively large value of $\omega_{ce}/\gamma_g$. Following that criterion, Figure \ref{fig:anisos}$b$ shows these thresholds using the same values of $f_e$ and $\Theta_{e,\parallel}$ used in Figure \ref{fig:anisos}$a$, but assuming $\gamma_g/\omega_{ce}=10^{-6}$ instead of $10^{-2}$. We see that for $\gamma_g/\omega_{ce}=10^{-6}$ the PEMW, PEMZ and OQES modes dominate for values of $f_e$ and $\Theta_{e,\parallel}$ very similar to the ones shown for $\gamma_g/\omega_{ce}=10^{-2}$. This implies that the role played by the different modes in controlling the electron temperature anisotropy in the different plasma regimes should not change significantly for realistic values of the shearing parameter $s$.\newline    

\noindent We now investigate whether increasing $\omega_{ce}^{\textrm{init}}/s$ affects the spectral evolution of the electrons. This is done in Figure \ref{fig:fldspec}$b$, where the final spectra ($t\cdot s=4$) are shown for runs with $\omega_{ce}^{\textrm{init}}/s=300, 600$, 1200 and 2400 (runs S300, S600, S1200 and S24000, respectively). No significant differences are seen in the final spectra, except for a slight hardening as $\omega_{ce}^{\textrm{init}}/s$ increases, which does not seem significant when comparing the cases $\omega_{ce}^{\textrm{init}}/s=$ 2400 and 1200. This shows that the magnetization parameter $\omega_{ce}^{\textrm{init}}/s$ does not play a significant role in the efficiency of the electron nonthermal acceleration.\newline

\begin{figure}[t!]  
\centering 
\includegraphics[width=8.8cm]{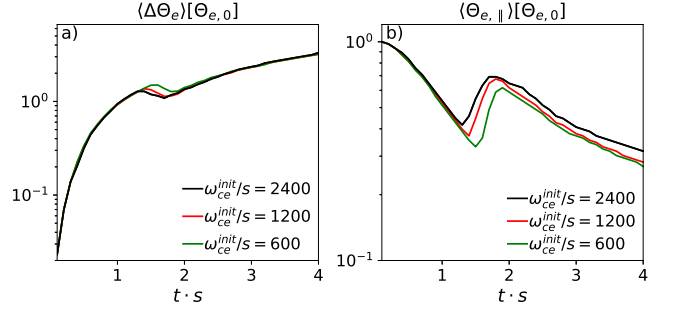}
\setlength{\abovecaptionskip}{-6pt plus 3pt minus 2pt}
\caption{Panel $a$: the evolution of the average $\Delta \Theta_e$ for the runs S2400 (black), S1200 (red) and S600 (green). Panel $b$: the evolution of the average $\Theta_{e,\parallel}$ for the same runs.} 
\label{fig:pressures} 
\end{figure}

\noindent The lack of dependence of the acceleration on $\omega_{ce}^{\textrm{init}}/s$ can be physically understood in terms of the relation between the effective pitch-angle scattering rate due to the instabilities, $\nu_{\textrm{eff}}$, and $s$. We estimate $\nu_{\textrm{eff}}$ using the evolution of $\Theta_{e,\parallel}$, which in a slowly evolving and homogeneous shearing plasma is given by \citep{SharmaEtAl2007}:
\begin{equation}
\frac{d\Theta_{e,\parallel}}{dt} + 2\Theta_{e,\parallel}\hat{\textbf{\textit{b}}}\hat{\textbf{\textit{b}}}: \nabla \textbf{\textit{v}} = \nu_{\textrm{eff}}\frac{2}{3}\Delta \Theta_e,
\label{eq:sharma1}
\end{equation}
where $\textbf{\textit{v}}$ is the plasma shear velocity, $\hat{\textbf{\textit{b}}}=\hat{\textbf{\textit{B}}}/B$ and $\Delta \Theta_e=\Theta_{e,\perp}-\Theta_{e,\parallel}$. Considering that $\textbf{\textit{v}}=-sx\hat{y}$, we can rewrite Equation \ref{eq:sharma1} as:
\begin{equation}
\frac{d\Theta_{e,\parallel}}{d(ts)} + 2\Theta_{e,\parallel}\hat{b}_x\hat{b}_y = \frac{2}{3}\Delta \Theta_e\frac{\nu_{\textrm{eff}}}{s}.
\label{eq:sharma2}
\end{equation}
Figure \ref{fig:pressures}$a$ shows the evolution of $\langle \Delta \Theta_e \rangle$ for the runs with $\omega_{ce}^{\textrm{init}}/s=600, 1200$ and 2400 (runs S600, S1200 and S2400). When $t\cdot s \gtrsim 2.2$, the factor $\Delta \Theta_e$ that appears on the right hand side of Equation \ref{eq:sharma2} is very similar in the three runs. The left hand side of Equation \ref{eq:sharma2} depends on the evolution of $\Theta_{e,\parallel}$, which, in the $t\cdot s \gtrsim 2.2$ regime, is also similar in the three runs as can be seen from Figure \ref{fig:pressures}$b$. This means that, for $t\cdot s \gtrsim 2.2$, the ratio $\nu_{\textrm{eff}}/s$ in Equation \ref{eq:sharma2} is fairly constant (within about $\sim 10\%$) for these three values of $\omega_{ce}^{\textrm{init}}/s$. Thus, for simulations with a fixed value of $\omega_{ce}^{\textrm{init}}$ but different $s$, the behaviors of $\langle \Delta \Theta_e \rangle$ and $\langle \Theta_{e,\parallel} \rangle$ for $t\cdot s \gtrsim 2.2$ imply that $\nu_{\textrm{eff}}$ should be approximately proportional to $s$. Because of that, on average, the number of times that the electrons are strongly deflected in a fixed interval of $t\cdot s$ should be largely independent of $\omega_{ce}^{\textrm{init}}/s$. Thus, in a stochastic acceleration scenario, we expect the acceleration effect during a fixed number of shear times ($s^{-1}$) to only depend on the dispersive properties of the unstable modes \citep[see, e.g.,][]{SummersEtAl1998}, which are not expected to depend on the value of $\omega_{ce}^{\textrm{init}}/s$. Indeed, as long as $\omega_{ce}^{\textrm{init}} \gg s$, the modes propagation and oscillations should occur rapidly (on timescales of $\sim \omega_{ce}^{-1}$), and should not be affected by the slowly evolving background (on timescales of $\sim s^{-1}$). These arguments thus imply that the acceleration efficiency should be largely independent of  $\omega_{ce}^{\textrm{init}}/s$. \newline

\noindent Notice that Figures \ref{fig:pressures}$a$ and \ref{fig:pressures}$b$ show that, in the $t\cdot s \lesssim 2.2$ regime, $\langle \Delta \Theta_e \rangle$ and $\langle \Theta_{e,\parallel} \rangle$ tend to depend more strongly on $\omega_{ce}^{\textrm{init}}/s$, so the electron acceleration efficiency should differ significantly during that period. However, Figure \ref{fig:fldspec}$a$ shows that for $t\cdot s \lesssim 2.2$ the dominant instabilities are mainly oblique modes, which is indicative of the dominance of OQES modes. Therefore, since no significant acceleration is expected to happen in that regime (as we showed in \S \ref{case1200}), the different evolutions of $\langle \Delta \Theta_e \rangle$ and $\langle \Theta_{e,\parallel} \rangle$ for $t\cdot s \lesssim 2.2$ should not have an appreciable effect on the final electron spectra. Although this analysis is valid for the rather limited range of values of $\omega_{ce}^{\textrm{init}}/s$ tested by our simulations, we expect the $\nu_{\textrm{eff}} \propto s$ relation to hold even for realistic values of $\omega_{ce}^{\textrm{init}}/s$. This is based on comparing the linear theory thresholds presented in Figures \ref{fig:anisos}$a$ and \ref{fig:anisos}$b$, which predict $\Delta \Theta_e/\Theta_{e,\parallel}$ to have essentially the same evolution (only differing by an overall factor $\sim 3$) when decreasing $\gamma_g/\omega_{ce}$ by four orders of magnitude.\newline
\begin{figure*}
\centering 
\hspace*{-.1cm}\includegraphics[width=16cm]{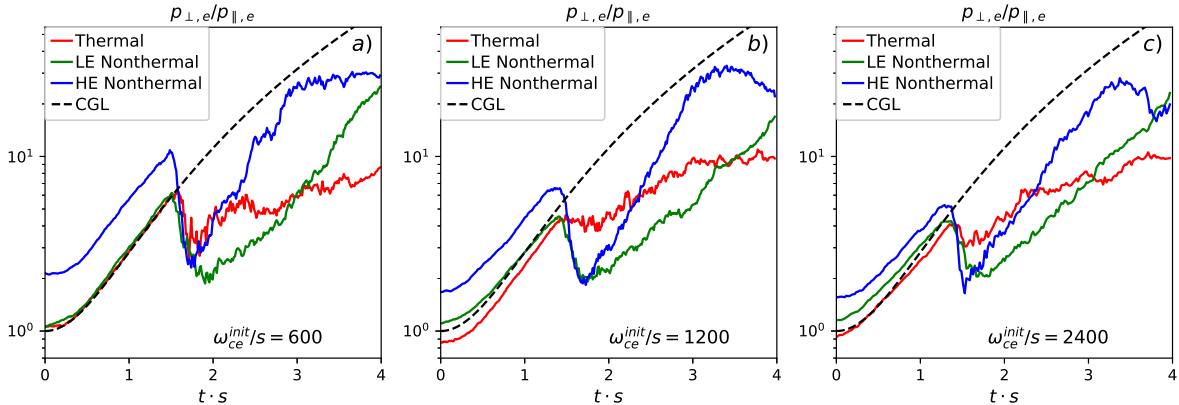}
\caption{Panels $a$, $b$ and $c$ show $p_{e,\perp}/p_{e,\parallel}$ for different populations in runs with $\omega_{ce}^{\textrm{init}}/s=600$, 1200 and 2400, respectively (runs S600, S1200 and S2400, respectively). The populations correspond to the thermal (red), low-energy nonthermal (green) and high-energy nonthermal (blue) electrons which are defined in \S \ref{case1200} according to their final energies. In the three panels we also show as dashed-black lines the estimated CGL evolution of $p_{e,\perp}/p_{e,\parallel}$ for an initially isotropic population.} 
\label{fig:tplowtemp_aniso} 
\end{figure*}
\subsection{Pitch-angle evolution}
\label{sec:pitchevol}

\noindent The evolution of the electron pitch-angle is important to determine the ability of the electrons to escape the flare looptops and precipitate towards the footpoints, which is a key ingredient for solar flare emission models \citep[e.g.,][]{MinoshimaEtAl2011}. In this section we investigate the way temperature anisotropy instabilities affect the pitch-angle evolution for electrons with different energies.\newline

\noindent As a measure of the average pitch-angle for different electron populations, Figure \ref{fig:tplowtemp_aniso}$b$ shows $p_{e,\perp}/p_{e,\parallel}$ for the thermal (red), low-energy nonthermal (green) and high-energy nonthermal (blue) electrons from run S1200. These three electron populations are defined according to their final energies, as we did in \S \ref{case1200} (in each of these cases, the pressures are calculated only considering the electrons in each population). For comparison, we also show as dashed-black lines the time dependence of $p_{e,\perp}/p_{e,\parallel}$ for a hypothetical, initially isotropic electron population that evolves according to the CGL equation of state. The three electron populations show evolutions similar to the CGL prediction until the onset of the OQES instability, which, as we saw in \S \ref{sec:visheating}, occurs at $t\cdot s \sim 1.4$. After that, the pitch-angle scattering tends to reduce $p_{e,\perp}/p_{e,\parallel}$ for the three populations, which occurs more abruptly for the low- and high-energy nonthermal electrons, whose pitch-angles by $t\cdot s \sim 1.7$ becomes $\sim 2-3$ times smaller than the ones of the thermal electrons. This shows that, even though the OQES modes (which dominate until  $t\cdot s \approx 2.2$) make a subdominant contribution to the nonthermal electron acceleration, they still have an important effect by significantly reducing the pitch-angle of the highest energy particles. After that, in the PEMZ dominated regime, the anisotropy evolves in the opposite way. In that case, the $p_{e,\perp}/p_{e,\parallel}$ ratios of both populations of nonthermal electrons grow more rapidly than for the thermal electrons. This is especially true for the high-energy nonthermal electrons, for which $p_{e,\perp}/p_{e,\parallel}$ reaches values $\sim 2-3$ times larger than for thermal electrons by the end of the run. \newline

\noindent This increase in the electron pitch-angle of the highest energy electrons due to PEMZ mode scattering is consistent with quasi-linear theory results that describe the stochastic acceleration of electrons by whistler/z modes in terms of the formation of a ``pancake" pitch-angle distribution for the most accelerated electrons \citep{SummersEtAl1998}. These results thus suggest that, assuming a more realistic solar flare scenario where the electrons were allowed to prescipitate towards the flare footpoints, the high-energy nonthermal electrons produced by PEMZ mode scattering should tend to be more confined to the looptop than the low-energy nonthermal and thermal electrons. \newline

\noindent Figures \ref{fig:tplowtemp_aniso}$a$ and \ref{fig:tplowtemp_aniso}$c$ show the same quantities as Figure \ref{fig:tplowtemp_aniso}$b$ but for runs with $\omega_{ce}^{\textrm{init}}/s=600$  and 2400 (runs S600 and S2400). After the triggering of the instabilities ($t\cdot s \sim 1.4$), there are no substantial differences between the three magnetizations, suggesting that the energy dependence of the pitch-angle evolution in our runs is fairly independent of $\omega_{ce}^{\textrm{init}}/s$.\newline

\begin{figure*}[t!]  
\centering 
\hspace*{-1cm}\includegraphics[width=20cm]{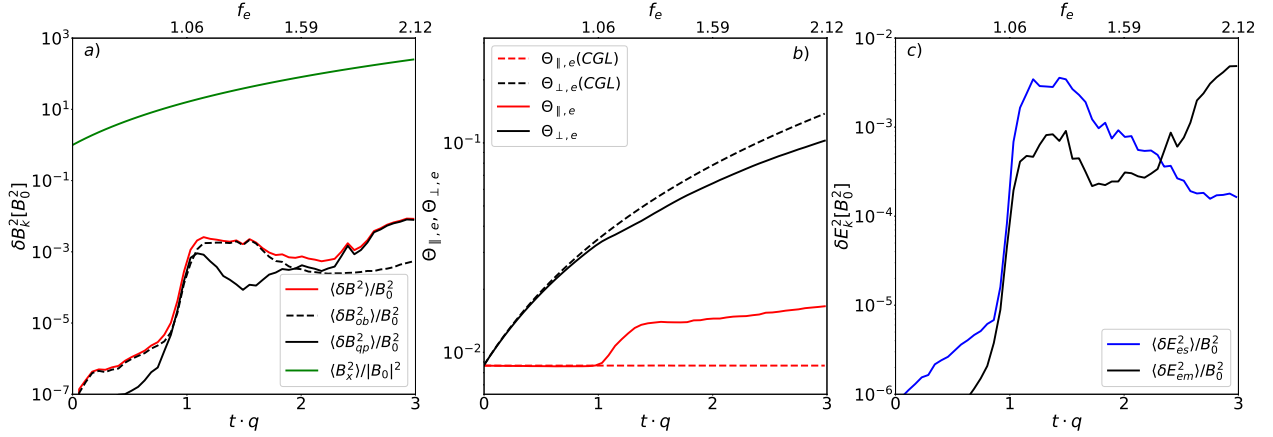}
\caption{We show fields and electron temperature evolutions for the compressing run C2400 ($\omega_{ce}^{\textrm{init}}/q=2400$) as a function of time $tq$ and of the instantaneous $f_e$. Panel $\it{a}$: in solid-green the evolution of the energy in the $x$ component of $\langle \textbf{\textit{B}}\rangle$. The solid-red line shows the energy in $\delta \textbf{\textit{B}}$, while the solid-black (dashed-black) line shows the contribution to the $\delta \textbf{\textit{B}}$ energy by the quasi-parallel (oblique) modes. Panel $\it{b}$: in solid-black (solid-red) the evolution of the electron temperature perpendicular (parallel) to $\langle \textbf{\textit{B}}\rangle$. The dashed-black (dashed-red) line shows the CGL prediction for the perpendicular (parallel) temperature. Panel $\it{c}$: in black (blue) the energy in the electromagnetic and electrostatic component of $\delta \textbf{\textit{E}}$.}
\label{fig:comp1} 
\end{figure*}

\subsection{Role of the initial conditions in the final electron energies}
\label{sec:initial}

\noindent An interesting aspect of the electron energy evolution is the correlation between the final electron energies and $i)$ their initial energies and $ii)$ their initial pitch-angles. The correlation between the initial and final energies can be seen from Figures \ref{fig:tplowtemp}$a$, \ref{fig:tplowtemp}$b$ and \ref{fig:tplowtemp}$c$, which show that for the thermal, low-energy nonthermal and high-energy nonthermal electrons, the initial energies are given by $\langle \gamma_e \rangle_0 -1 \sim 0.01, \sim 0.025$ and $\sim 0.04$, respectively. This energy correlation is also expected from the quasi-linear  theory  results of \cite{SummersEtAl1998}, which show that the maximum energy that electrons can acquire due to stochastic acceleration by whistler/z modes increases for larger initial energies (assuming a fixed value of $f_e$).\newline

\noindent Additionally, Figure \ref{fig:tplowtemp_aniso}$b$ shows that the initial values of $p_{e,\perp}/p_{e,\parallel}$ are larger for electrons that end up being more energetic, with the initial $p_{e,\perp}/p_{e,\parallel}$ being $\sim 0.9$, $\sim 1.1$ and $\sim 1.7$ for the thermal, low-energy nonthermal and high-energy nonthermal electrons, respectively. 
This is consistent with the fact that, before the PEMZ dominated regime ($t\cdot s \lesssim 2.2$), the three electrons populations gain their energy mainly due to AV, as can be seen from the three panels in Figure \ref{fig:tplowtemp}. This means that during that period, energy is gained more efficiently by electron populations with larger average pitch-angle. After that, given that the scattering by OQES waves strongly reduces the pitch-angle of the nonthermal electrons, their energy gain is no longer related to their $p_{e,\perp}/p_{e,\parallel}$, but to the more efficient acceleration that the PEMZ modes cause on initially more energetic electrons.\newline

\section{Comparison with the compressing case}
\label{sec:compressing}
In this section we show the effect of driving the $T_{e,\perp} > T_{e,\parallel}$ anisotropy by using compressive instead of shearing simulations. In order to ease comparison, the compressing simulations use the same initial conditions as in the shearing case ($f_e \approx 0.53$, $\Theta_e^{\textrm{init}}=0.00875$) and are also run until $f_e$ reaches $f_e \approx 2$. This way we test whether the transition between the regimes dominated by OQES and PEMZ modes also occurs in the compressing runs, and whether this is also accompanied by an increase in the acceleration efficiency. Since in the compressing runs, $\textbf{\textit{v}} = -q(y\hat{y}+z\hat{z})/(1+qt)$, magnetic flux freezing makes the mean magnetic field grow by the same factor as the mean electron density: $(1+qt)^2$. Thus, in order to make $f_e$ grow from 0.53 to $\sim 2$, our compressing simulations run until $tq=3$. Similarly to our analysis for the shearing simulations, first we describe the interplay between the evolution of the mean magnetic field and the temperature anisotropy. After that we concentrate on the nonthermal evolution of the electron energy spectrum. We show that, although in the compressing case there is significant acceleration driven by the PEMZ modes, this acceleration is less efficient that in the shearing case.\newline 

\subsection{Anisotropy regulation in the compressing case}
\label{sec:interplay_compressing}

Figure \ref{fig:comp1}$a$ shows in solid-green the growth of the $x$ component of $\langle \textbf{\textit{B}}\rangle $ for simulation C2400 ($\omega_{ce}^{\textrm{init}}/q=2400$), which evolves as $|B_x| = |\langle \textbf{\textit{B}}\rangle | = B_0(1+qt)^2$. Because of this magnetic amplification, initially $\Theta_{e,\perp}$ and $\Theta_{e,\parallel}$ grow and stay constant, respectively, as seen from the solid-black and solid-red lines in Figure \ref{fig:comp1}$b$. For $t\cdot q \lesssim 1$, $\Theta_{e,\perp}$ and $\Theta_{e,\parallel}$ coincide with the adiabatic CGL evolution shown by the dashed-black and dashed-red lines, respectively. The departure from the CGL behavior at $t\cdot q \approx 1$ is coincident with the growth and saturation of $\delta \textbf{\textit{B}}$, shown by the solid-red line in Figure \ref{fig:comp1}$a$. This shows that $\Theta_{e,\perp}$ and $\Theta_{e,\parallel}$ are regulated by the pitch-angle scattering provided by temperature anisotropy unstable modes, as it is the case for the shearing simulations shown in \S \ref{sec:visheating}.\newline 

\begin{figure}[t!]  
\centering 
\hspace*{-0.8cm}\includegraphics[width=9.8cm]{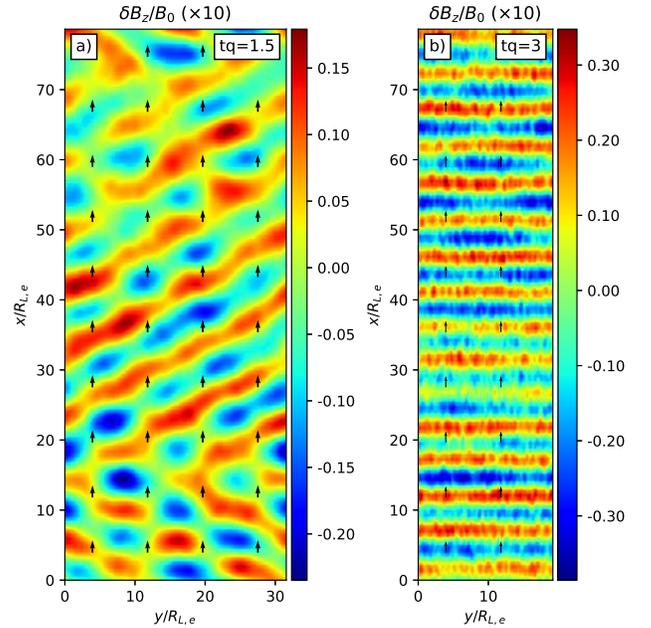}
\setlength{\abovecaptionskip}{-6pt plus 3pt minus 2pt}
\caption{For run C2400 ($\omega_{ce}^{\textrm{init}}/q=2400$) we show the 2D distribution of $\delta \textit{B}_z$ at $t\cdot q = 1.5$ (panel $a$) and $t\cdot q = 3$ (panel $b$), normalized by $B_0$. The black arrows show the direction of $\langle \textbf{\textit{B}} \rangle$.} 
\label{fig:fldcomp_bz} 
\end{figure}

\noindent Figures \ref{fig:fldcomp_bz}$a$ and \ref{fig:fldcomp_bz}$b$ show a 2D view of $\delta \textit{B}_z$ for the unstable modes in run C2400 at $t\cdot q = 1.5$ (after the saturation of $\delta \textbf{\textit{B}}$) and $t\cdot q = 3$, respectively. Although the box has initially a square shape, the effect of compression makes its $y$-size decreases with time as $1/(1+qt)$, which explains the progressively more elongated shape of the box shown by the two panels. The dominant modes at $t\cdot q = 1.5$ have wavevectors that are oblique with respect to the direction of the mean magnetic field $\langle \textbf{\textit{B}} \rangle$, which is shown by the black arrows. At $t\cdot q = 3$ the waves propagate mainly along $\langle \textbf{\textit{B}}\rangle$, which shows that by the end of the simulation the modes are quasi-parallel. The time for the transition between the dominance of oblique to quasi-parallel modes can be obtained by calculating the magnetic energy contained in each type of mode. As for the shearing simulations shown in \S \ref{sec:visheating}, we define the modes as oblique (quasi-parallel) when the angle between their wave vector $\textbf{\textit{k}}$ and $\langle\textbf{\textit{B}}\rangle$ is larger (smaller) than 20$^{\circ}$. Figure \ref{fig:comp1}$a$ shows the magnetic energies in oblique and quasi-parallel modes using dashed-black and solid-black lines, which are denoted by $\langle \delta B_{ob}^2 \rangle$ and $\langle \delta B_{qp}^2 \rangle$, respectively. The oblique modes dominate until $t\cdot s \approx 2$, which corresponds to $f_e \approx 1.6$. After that, the energy in quasi-parallel modes dominates the magnetic fluctuations. The existence of these oblique and quasi-parallel regimes, and the fact that the transition occurs when $f_e \approx 1.6$, implies that dominance of the different modes is essentially determined by the value of $f_e$, as it was found in the case of the shearing simulations. \newline

\begin{figure}[t!]  
\centering 
\hspace*{-.8cm}\includegraphics[width=9.8cm]{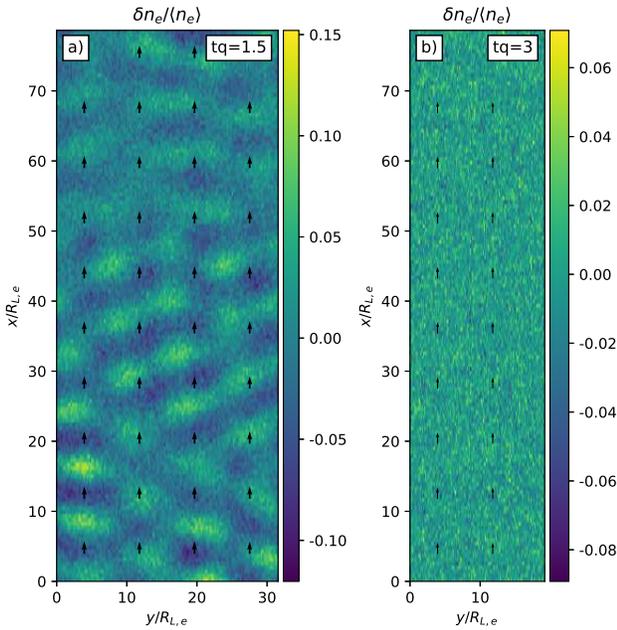}
\setlength{\abovecaptionskip}{-6pt plus 3pt minus 2pt}
\caption{Same as Fig. \ref{fig:fldcomp_bz} but for the electron density fluctuations $\delta n_e$, normalized by the average density $\langle n_e \rangle$.} 
\label{fig:fldcomp_dens} 
\end{figure}

\noindent Similarly to what occurs in the shearing case, the oblique modes in the compressing runs show significant fluctuations in the electron density $n_e$. This is shown in Figure \ref{fig:fldcomp_dens}$a$, which shows $\delta n_e$ at $t\cdot q=1.5$. The fluctuations in the electron density practically disappear when the quasi-parallel modes dominate, as shown by Figure \ref{fig:fldcomp_dens}$b$, which shows $\delta n_e$ at $t\cdot q=3$. These different behaviors of $\delta n_e$ suggest the existence of a significant electrostatic component in the electric field $\delta \textbf{\textit{E}}$ when the oblique modes dominate. This is confirmed by Figure \ref{fig:comp1}$c$, which shows the energy contained in the electric field fluctuations $\delta \textbf{\textit{E}}$ as a function of time, separating it into electrostatic ($\delta E^2_{es}$) and electromagnetic ($\delta E^2_{em}$) components, which are shown in blue and black lines, respectively. When the oblique modes dominate ($t\cdot q \lesssim 2$) the electric field energy is dominated by $\delta E^2_{es}$; after that it gradually becomes dominated by $\delta E^2_{em}$. Figure \ref{fig:comp1}$c$ also shows that this transition occurs when the instantaneous $f_e\sim 1.6-1.7$ (the instantaneous $f_e$ is shown by the upper horizontal axis), which is fairly similar to the transition seen in the case of shearing simulations at $f_e\sim 1.2-1.5$.\newline
\begin{figure*}[t!]  
\centering 
\hspace*{-0.2cm}\includegraphics[width=18.4cm]{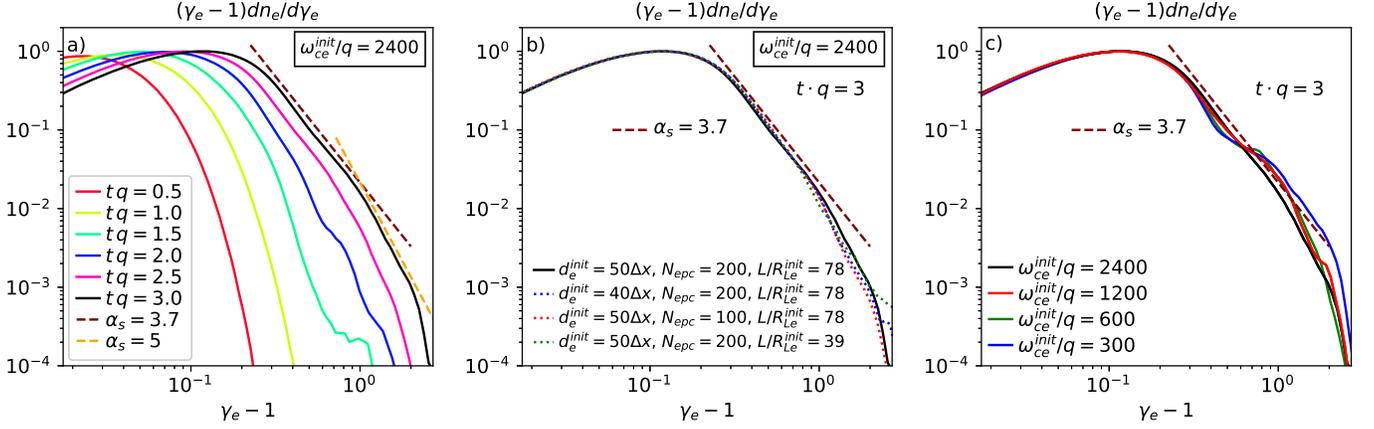}
\setlength{\abovecaptionskip}{-6pt plus 3pt minus 2pt}
\caption{Panel $a$: the electron energy spectrum for run C2400 for different values of $t\cdot q$, where $\gamma_e$ is the electron Lorentz factor.  The dashed-brown line shows a power-law of index $\alpha_s \approx 3.7$, which at $t\cdot q \approx 2.5-3$ resembles the nonthermal tail up to a break at $\gamma_e \sim 1$. \textit{Panel b:} test of numerical convergence of the final spectrum ($t\cdot q =3$). We compare run C2400 (solid black; $d_e^{\textrm{init}}/\Delta x=50$, $N_{epc}=200$, $L/R_{Le}^{\textrm{init}}=78$) with runs using $i)$ a smaller time and space resolution $d_e^{\textrm{init}}/\Delta x=40$ (run C2400a, in blue-dotted line), $ii)$ a smaller $N_{epc}=100$ (run C2400b, in red-dotted line) and $iii)$ a smaller box size $L/R_{Le}^{\textrm{init}}=39$ (run C2400c, in green-dotted line). No significant difference can be seen between the different spectra, except for the highest resolution run C2400 having a slightly harder spectrum. \textit{Panel c:} the final electron  spectra  for  runs  C2400  (black),  C1200  (red),  C600(green) and C300 (blue).} 
\label{fig:fldspec_comp} 
\end{figure*}

\subsection{Electron acceleration in the compressing case}
\label{sec:acceleration_compressing}

\noindent As in the shearing case, PEMZ modes driven by plasma compression also accelerate electrons, although there are some significant differences. Figure \ref{fig:fldspec_comp}$a$ shows $dn_e/d\ln(\gamma_e-1)$ for run C2400 at different times $t\cdot q$. We see the rapid growth of a nonthermal tail starting at $t\cdot q\approx 2$. After $t\cdot s\approx 2.5$ this tail can be approximated as a power-law of index $\alpha_s \approx 3.7$ with a break at $\gamma_e-1\sim 1$, where the spectrum becomes significantly steeper. The fact that most of the nonthermal behavior starts when the PEMZ modes become dominant ($t\cdot q\sim 2$), suggests that, as in the shearing case, the nonthermal acceleration is mainly driven by the PEMZ modes.\newline

\noindent One significant difference between the shearing and compressing case is that the latter gives rise to a softer nonthermal component in the final electron energy spectrum than the former (a power-law of index $\alpha_s \approx 3.7$ instead of $\alpha_s \approx 2.9$). This difference is not surprising given that the overall electron energy gain when $f_e$ evolves from $f_e=0.53$ to $f_e\approx 2$ is significantly larger for the compressing runs. While in the shearing case the final value of $\Theta_{e,\perp}$ is about $\sim 3$ times larger than its initial value (see the solid-black line in Figure \ref{fig:sf1}$b$), in the compressing case the final $\Theta_{e,\perp}$ is about $\sim 10$ times larger than the initial $\Theta_{e,\perp}$ (see the solid-black line in Figure \ref{fig:comp1}$b$). Since at the end of both types of simulations $\Theta_{e,\perp} \gg \Theta_{e,\parallel}$, this difference implies that by the end of the compressing runs the electrons are significantly hotter than in the shearing runs. The different electron internal energies may affect the efficiency with which electrons gain energy from their interaction with the PEMZ modes, as it has been shown by previous quasi-linear theory studies of the stochastic acceleration of electrons by whistler/z modes \citep{SummersEtAl1998}. This implies that shearing and compressing runs that start with the same electron parameters should not necessarily produce equally hard nonthermal component in the electron spectrum after $f_e$ has increased from 0.53 to $\sim 2$.\newline 

\noindent Figure \ref{fig:fldspec_comp}$b$ shows a numerical convergence test of the final spectrum at $t\cdot q=3$. It compares run C2400 ($d_e^{\textrm{init}}/\Delta x=50$, $N_{epc}=200$ and $L/R_{Le}^{\textrm{init}}=78$ as shown in Table \ref{table27}) with a run with $d_e^{\textrm{init}}/\Delta x=40$ (run C2400a, in blue-dotted line), a run with $N_{epc}=100$ (run C2400b, in red-dotted line) and a run with $L/R_{Le}^{\textrm{init}}=39$ (run C2400c, in green-dotted line). No significant difference can be seen between the different spectra, with only a slight hardening for the highest resolution run C2400, which implies that our results are reasonably well converged numerically.\newline

\noindent The effect of varying $\omega_{ce}^{\textrm{init}}/q$ is investigated in Figure \ref{fig:fldspec_comp}$c$, where the final spectra ($t\cdot q=3$) are shown for runs with $\omega_{ce}^{\textrm{init}}/q=300, 600$, 1200 and 2400 (runs C300, C600, C1200 and C2400, respectively). As $\omega_{ce}^{\textrm{init}}/q$ increases there is a hardening of the low energy part of the nonthermal tail (0.3 $\lesssim \gamma_e-1 \lesssim 0.8$), which converges towards a power-law tail of index $\alpha_s \approx 3.7$ with little difference between the cases with $\omega_{ce}^{\textrm{init}}/q=1200$ and 2400. Additionally, as $\omega_{ce}^{\textrm{init}}/q$ increases, there is a decrease in the prominence of a high energy bump at $\gamma_e-1 \sim 1$, which essentially disappears for $\omega_{ce}^{\textrm{init}}/q=2400$. This suggests that the magnetizations $\omega_{ce}^{\textrm{init}}/q$ used in our compressing runs provides a reasonable approximation to the expected spectral behavior in realistic flare environments.\newline

\section{Possible role of collisions}
\label{sec:collisions}

\noindent In this work we have assumed a negligible role of Coulomb collissions between electrons. We validate this assumption by calculating the Coulomb collision rate $\nu_{ee}$ for electrons with temperature $T_e$ and density $n_e$ \citep{Spitzer1962}:
\begin{equation}
    \nu_{ee} \approx 0.2 \textrm{ sec}^{-1} \Big(\frac{n_e}{10^{9}\textrm{cm}^{-3}}\Big)\Big(\frac{T_e}{50 \textrm{MK}}\Big)^{-3/2},
\label{eq:col}
\end{equation}
where we have assumed a Coulomb logarithm of 20 (appropriate for $n_e\approx 10^{9}\textrm{cm}^{-3}$ and $T_e \approx$ 50 MK). Since, for our fiducial parameters $n_e\approx 10^{9}\textrm{cm}^{-3}$ and $T_e \approx$ 50 MK, $\nu_{ee}$ is significantly smaller than the rate at which temperature anisotropy growth would be driven in contracting looptops ($s \sim 1$ sec$^{-1}$ for $B \sim 100$ G, as we estimated in \S \ref{sec:realistic}), our collisionless approach is valid in the low density cases ($n_e\sim 10^{8}-10^{9}\textrm{cm}^{-3}$). However, Equation \ref{eq:col} shows that, for $n_e \gtrsim 10^{10}$ cm$^{-3}$, collisions could become dominant, probably reducing the efficiency of the electron acceleration.\newline   

\section{Summary and conclusions}
\label{sec:conclu}

\noindent Using 2D particle-in-cell plasma simulations we study the effect of temperature anisotropy instabilities on electron acceleration under conditions suitable for ALT regions in solar flares. In our simulations we drove the growth of a $T_{e,\perp} > T_{e,\parallel}$ anisotropy using the adiabatic invariance of the electron magnetic moment $\mu_e$ in a growing magnetic field $B$, which is achieved by imposing either a shearing or a compressing plasma motion. In both cases, when the difference between $T_{e,\perp}$ and $T_{e,\parallel}$ is large enough, different plasma modes become unstable and, through pitch-angle scattering, limit the anisotropy growth. Since $B$ continuously grows in our simulations, our setup drives the instabilities into their nonlinear, saturated regime, allowing the $T_{e,\perp}$ and $T_{e,\parallel}$ anisotropy to self-regulate and capturing the long-term effect of the instabilities on the electron spectra. \newline

\noindent Our study considers an initial electron temperature $T_e\approx 52$  MK and an electron density and magnetic field $B$ such that $f_e$ evolves from $f_e \approx 0.53$ to $\approx 2$. Our results are summarized as follows: 

\begin{enumerate}
\item Both in the shearing and compressing runs, electrons are efficiently accelerated mainly by the inelastic scattering provided by unstable PEMZ modes, which dominate for $f_e \gtrsim 1.2-1.7$. This acceleration corresponds to a transfer of energy from the electrons in the thermal part of the spectrum to the electrons in the nonthermal tail, with the PEMZ modes playing the role of carriers of that energy. When $f_e \lesssim 1.2-1.7$, pitch-angle scattering is mainly provided by OQES modes and the nonthermal acceleration is rather inefficient.\newline

\item By the end of the shearing runs, the spectrum contains a nonthermal tail that can be approximated as a power-law of index $\alpha_s \approx 2.9$, in addition to a high energy bump that reaches energies of $\sim 300$  keV. By the end of the compressing runs, the spectrum has an approximate power-law tail of index $\alpha_s \approx 3.7$, with a break at $\sim 500$ keV; at higher energies, $\alpha_s \sim 5$ (as shown by the dashed-orange line in Figure \ref{fig:fldspec_comp}$a$). This difference between the shearing and compressing runs is as expected given the different evolution of electron temperatures in these two types of runs.\newline

\item Our results are largely independent of the ratios $\omega_{ce}/s$ or $\omega_{ce}/q$, when these ratios are sufficiently large. This implies that our study can be extrapolated to realistic solar flare conditions, where $\omega_{ce}/s$ or $\omega_{ce}/q$ should be several orders of magnitude larger than in our simulations.  
\end{enumerate}

\noindent In conclusion, our simulations show that, under conditions expected in ALT sources, electron temperature anisotropy instabilities have the potential to contribute to the acceleration of electrons, probably as a complement to the acceleration processes expected in reconnection current sheets in solar flares. Interestingly, the spectral index observed from our shearing runs ($\alpha_s \approx 2.9$) is within the range of inferred indices in some ALT sources \citep[e.g.,][]{AlexanderEtAl1997}. Also, the spectral index obtained from our compressing runs ($\alpha_s \approx 3.7$) agrees reasonably well with the spectral index $\alpha_s \approx 3.6$ inferred from multiwavelenth observations of the ALT source in the X8.2-class solar flare of September 10, 2017 \citep{ChenEtAl2021}. In this flare, the power-law tail shows a break at $\sim 160$ keV, and, at higher energies, $\alpha_s \sim 6$. This feature is qualitatively similar to the spectral break down observed in our compressing runs, although in our simulations the break occurs at $\sim 500$ keV, with $\alpha_s \sim 5$ at higher energies.\newline 

\noindent We point out, however, that the present study is not intended to make precise predictions regarding observationally inferred nonthermal electron tails in ALT sources. Indeed, the anisotropy driving implemented in this work constitutes a simplified, local model for the way the electron velocity distribution may evolve in ALT sources. A global description should consider the electrons ability to escape the looptop region \citep[which may give rise to a loss-cone velocity distribution with $T_{e,\perp} > T_{e,\parallel}$; e.g.][]{FleishmanEtAl1998}, as well as a more realistic prescription for the magnetic field evolution. Also, the initial values of $f_e$ and $T_e$ chosen in this work, although appropriate for ALT sources, do not represent the whole range of possible conditions in these environments. For these reasons, we consider this study a first step in assessing the possible role of electron temperature anisotropy instabilities in accelerating electrons in solar flares, under specific initial conditions and assuming that this anisotropy is driven by either shearing or compressing plasma motions. We defer the study of the effect of a wider range of initial plasma conditions as well as of the global loop dynamics to future works.\newline 


\acknowledgements
\noindent MR thanks support from a Fondecyt Regular Grant No. 1191673. AO acknowledges support from a Beca EPEC-FCFM. DV is supported by the STFC Ernest Rutherford Fellowship ST/P003826/1 and STFC Consolidated Grant ST/S000240/1. This research was supported in part by the National Science Foundation under Grant No. NSF PHY-1748958. Most of the numerical simulations included in this work were performed at the National Laboratory for High Performance Computing (NLHPC) of the Center for Mathematical Modeling of University of Chile (ECM-02).

\end{document}